\documentclass[prb,aps,twocolumn,epsfig,rotate,showpacs]{revtex4-1}
\usepackage{epsfig}
\usepackage{graphicx}
\usepackage{amsmath}
\usepackage{amssymb}

\newcommand{\beq}{\begin{eqnarray}}
\newcommand{\eeq}{\end{eqnarray}}
\begin{document}
\title{Validity of the Wiedemann-Franz law in small molecular wires}
\author{Vinitha Balachandran}
\affiliation{ Center for Nonlinear and Complex Systems,
Universit\`a degli Studi dell'Insubria, Via Valleggio 11, 22100 Como, Italy}
\author{Riccardo Bosisio}
\email{Present address:
Service de Physique de l'\'Etat Condens\'e (CNRS URA 2464),
IRAMIS/SPEC, CEA Saclay, 91191 Gif-sur-Yvette, France.
}
\affiliation{ Center for Nonlinear and Complex Systems,
Universit\`a degli Studi dell'Insubria, Via Valleggio 11, 22100 Como, Italy}
\author{Giuliano Benenti}
\affiliation{CNISM \& Center for Nonlinear and Complex Systems,
Universit\`a degli Studi dell'Insubria, Via Valleggio 11, 22100 Como, Italy}
\affiliation{Istituto Nazionale di Fisica Nucleare, Sezione di Milano,
via Celoria 16, 20133 Milano, Italy}

\begin{abstract}
We report our investigations on the finite-size effects of the Lorenz number in a molecular wire. Using Landauer-B\"uttiker formalism, we find that for sufficiently long wires
there are two validity regimes of the Wiedemann-Franz (WF) law, the cotunneling and
the sequential tunneling regimes, while in small systems only the first
regime survives.
We compare our results with the standard Kubo formalism and explain its failure to obtain the WF law in small systems. Furthermore, our studies on exponentially localized disordered wires show that the Lorenz number
value $\mathfrak{L}_{0}=(\pi^{2}/3)(k_B/e)^2$ predicted by the WF law is obtained only in the cotunneling regime.
Also, the Lorenz number $\mathfrak{L}$ exhibits a typical distribution at different temperatures corresponding to different tunneling processes. In particular, first-order tunneling results in a low value of $\mathfrak{L}$ whereas second-order tunneling recovers the universal value $\mathfrak{L}_{0}$.
\end{abstract}
%73.23.-b Electronic transport in mesoscopic systems
%73.63.Nm Quantum wires
\date{\today}
\maketitle

\section{introduction}
Thermoelectricity is attracting huge scientific interest on account of its applications in future energy production and utilization.
\cite{te,Majumdar,dresselhaus,snyder,shakuori,BC11}
Much of the work has focused on developing efficient thermoelectric materials to convert waste heat energy to electric current and, in reverse, to perform refrigeration. Compared to bulk materials, low-dimensional systems have the potential to achieve
improved thermoelectric efficiency owing to their highly peaked density of states
and to the high density of interfaces that could be used to reduce parasitic
heat flow.~\cite{ztone,heremans,datta,linke}
In this regard, studies related to thermal and electrical transport in nano scaled molecular wires have gained considerable attention.

Thermoelectric conversion efficiency is characterized by the figure of merit $ZT =T S^{2}\sigma/\kappa $, where $S$ is the thermopower and $\sigma$ and $\kappa$ are the electrical and thermal conductivity at temperature $T$. For practical applications, it is required that $ZT\gg1$. However, this is practically hard to achieve, as the thermal and electrical conductivity are related by the Wiedemann-Franz (WF) law, which states
that the Lorenz number
$\mathfrak{L}=\kappa/\sigma T$ is constant. \cite{WF} The constant value
$\mathfrak{L}_{0}=(\pi^{2}/3)(k_B/e)^2$ for non-interacting systems, where
$k_B$ is the Boltzmann constant and $e$ the charge carried by each electron. The
WF law follows from the single-particle Fermi liquid (FL) theory which assumes that both electric and thermal current are carried by the same FL particles at sufficiently low temperatures, so that the Sommerfeld expansion can be
applied.~\cite{asc} The WF law is valid in the thermodynamic limit of a non-interacting system even in the presence of arbitrary disorder provided that the FL theory holds. \cite{WFdisnon} Studies in interacting systems showed that the law is  violated largely due to the non-FL behavior. \cite{WFint1,WFint3,WFint4,WFint2,WFint5}

However, studies reported mainly so far have focused on either one or two quantum molecules or on the thermodynamic limit. In mesoscopic physics, the thermodynamic limit is meaningless and one is interested in the transport properties of finite systems. Finite-size effects are expected to influence the properties considerably in low-dimensional quantum systems. For instance, using the Landauer-B\"uttiker formalism, Vavilov and Stone showed that deviation from the WF law occurs around the Thouless temperature $T_{c}$, where the FL theory still holds. \cite{viol}  Note however, that these investigations were carried out on diffusive mesoscopic conductors.  Studies of finite-size effects on one-dimensional integrable systems showed that the thermopower $S$ does not follow the relation $S\propto T$ as expected in the thermodynamic limit. \cite{Sfinite} Furthermore, the Lorenz number is shown to diverge in the infinite-frequency limit of a finite-sized closed system. \cite{Lshastry} The above-mentioned works follow from the Kubo formula, without explicitly considering the connections to the baths. Recently, the transport properties obtained using the Kubo formula were shown to differ significantly from the Redfield quantum master equation approach (QME), which takes into account the effect of baths. \cite{divkubo} In practice, it is required that both
ends of the molecular wire are in contact with baths, which are able to exchange
charges and energy with the wire. The validity regimes of the WF law in these finite sized clean molecular wires is still an open question.

Following the seminal work of Anderson, \cite{Anderson} disorder has played an important role in understanding the transport properties of mesoscopic systems. Moreover, disorder has interesting effects on the properties of a finite system as the transport properties are affected considerably, depending upon the system's size. For instance, a disordered system shows ballistic transport if the localization length is very large compared to the system length. \cite{conduc}  Conductivities of strongly localized systems decrease exponentially with the system size. \cite{expdec} However, it is not yet clear how the Lorenz number varies with disorder in finite systems. When considering disordered systems, it is worthwhile to study the statistical distribution of observables. Studies in theses directions showed that the conductances of a strongly disordered system follow a log-normal distribution. \cite{pichard} Hence, it is interesting to explore whether the Lorenz number still holds this log-normal distribution.

In this paper, we investigate the validity regimes of the WF law in a finite
non-interacting molecular wire attached to reservoirs using the 
Landauer-B\"uttiker formalism for phase-coherent quantum transport
from one reservoir to the other.
With this approach we obtain conductances
rather than conductivities and the Lorenz number is redefined as
$\mathfrak{L}=\Xi/G T$, with $G$ and $\Xi$ the electrical and thermal
conductances, respectively. We compare our results with two other commonly used formalisms, namely the standard Kubo formalism and the QME, in exploring the transport properties of finite-sized systems. In the latter part of the work, the studies are extended to disordered systems.  In particular, our results show that the validity regimes of the WF law depend on the finite size of the system even for a non-interacting system. For long clean wires, there are saturation plateaus of Lorenz number at the universal value of $(\pi^{2}/3)(k_B/e)^2$ in two temperature regimes, corresponding to the cotunneling and the sequential tunneling process. As the wire length is reduced, one of the plateaus vanishes. The validity of the law in different tunneling regimes is explained in terms of the energy integrals giving the conductances following the studies of Vavilov and Stone in Ref. \cite{viol}. We also find that the standard Kubo formalism and QME approach fail, for any system size, to obtain WF law in the cotunneling regime. With disordered localized wires, WF law is valid only in the cotunneling regime.  In addition, Lorenz number shows typical distribution at different temperature regimes.

The outline of the paper is as follows. We introduce the model system and discuss the Landauer-B\"uttiker formalism to calculate conductances and the Lorenz number
in Sec.~\ref{sec:Model and Formalism}. Numerical results are discussed for a clean model  in Sec.~\ref{sec:Clean Wire}  and for a disordered one in Sec.~\ref{sec:Disordered Wire}. Finally, we draw our conclusions in Sec.~\ref{sec:Conclusions}.

\section{Model and Formalism}
\label{sec:Model and Formalism}

We consider a molecular junction formed by connecting a one-dimensional molecular wire between two electrodes. The Hamiltonian of our molecular system is
\begin{eqnarray}\label{ham}
{H}&=&{H}_{W}+{H}_{E}+{H}_{WE}; \nonumber\\
    {H}_{W}&=&-t\sum_{i=1}^{N-1}(c_{i}^{\dag}c_{i+1}+{\rm h.c.}), \nonumber\\ {H}_{E}&=&\sum_{j=L,R}\sum_{k}E_{kj}d_{kj}^{\dag}d_{kj},
     \nonumber\\  {H}_{WE}&=& \sum_{k}(t_{kL}c_{1}^{\dag}d_{kL}+t_{kR}c_{N}^{\dag}d_{kR}+{\rm h.c.}).
\end{eqnarray}
The first term ${H}_{W}$ corresponds to a wire of $N$ sites with nearest-neighbor interactions, the second term ${H}_{E}$, to the two electrodes left ($L$) and right ($R$); and the last term ${H}_{WE}$, to the wire-electrode coupling.  The operators $c_{i}^{\dag}(c_{i})$ and $d_{kj}^{\dag}(d_{kj})$ are creation (annihilation) operators of electrons in the wire and electrode $j$, respectively. $t$ is the hopping constant, $E_{kj}$ is the energy of the $k$th electron in the $j$th electrode and $t_{kj}$ is its tunneling amplitude.  Here, the electrodes are reservoirs of non-interacting electrons in equilibrium at some temperature $T$  and
electrochemical potential $\mu$.

Below we outline the calculations of thermoelectric properties in our model using the 
Landauer-B\"uttiker formalism. The current through the wire is due to the electrons tunneling from one electrode to another. An electron at a given energy $E$ scatters at the junction and can be transmitted through it or reflected back. The probability of tunneling across the junction is given by the transmission coefficient $\tau(E)$.  Hence, the electric ($J_{e}$) and thermal ($J_{q}$) currents (from left to right
reservoir) in the molecular wire are given by \cite{Landauer}
\begin{eqnarray}\label{current}
  J_{e} &=& \frac{e}{h}\int dE \tau(E)[f_{L}(E)-f_{R}(E)],\nonumber\\
   J_{q} &=& \frac{1}{h} \int dE (E-\mu) \tau(E)[f_{L}(E)-f_{R}(E)].
\end{eqnarray}
Here,  $e$ is the electronic charge, $h$ is Planck's constant, and $f_{L}(E)$ and $f_{R}(E)$ are the electron Fermi distributions in the left ($L$) and right ($R$) electrodes with temperature $T_{L,R}$ and electrochemical potential $\mu_{L,R}$. ($f_{L,R}(E)=\{\mathrm{exp}[(E-\mu_{L,R})/k_{B}T_{L,R}]+1\}^{-1}$ ,where $k_B$ is the Boltzmann constant.)
In this work, we are interested in the linear response of the system and hence assume that the differences $\Delta\mu=\mu_{L}-\mu_{R}$ and $\Delta T=T_{L}-T_{R}$ are infinitesimally small. Hence, in Eq. (\ref{current}), the electrochemical potential $\mu\approx\mu_{L}\approx \mu_R$ and the temperature $T\approx T_{L}\approx T_R$.

Using the nonequilibrium Green's function technique, the transmission coefficient can be expressed as,
\begin{equation}\label{tf}
    \tau(E)=\mathrm{Tr}(\Gamma_{L}(E)G_{s}^{\dag}(E)\Gamma_{R}(E)G_{s}(E)),
\end{equation}
where $\mathrm{Tr}$ is the trace, $G_{s}(E)=(E-H_{W}-\Sigma_{L}-\Sigma_{R})^{-1}$ is the retarded single-particle Green's function operator, and $\Gamma_{L,R}(E)=i[\Sigma_{L,R}(E)-\Sigma_{L,R}^\dagger(E)]$  are the level broadening functions. $\Sigma_{L}$ and $\Sigma_{R}$  are the retarded self-energies of the left and right electrodes, respectively.

We assume a wide band limit of the electrodes. Hence the level widths are energy independent and are given by $\gamma_{j}=2\pi\sum_{k}|t_{kj}|^{2}\delta(E-E_{kj})$. Furthermore, we take $\gamma_{L}=\gamma_{R}=\gamma$. Thus, $\Gamma_{L}=\gamma c_1^\dagger c_1$,
$\Gamma_{R}=\gamma c_N^\dagger c_N$ and  Eq. (\ref{tf}) can be rewritten as
\begin{equation}\label{tf1}
   \tau(E)=\gamma^{2}|\langle 1|G_{s}(E)|N\rangle|^{2}.
\end{equation}
Note that $\gamma$ is the coupling strength, which physically measures the rate at which the electrons tunnel across the junction.

Using the Taylor expansion,
\begin{equation}\label{tf2}
    f_{L}(E)\approx
f_{R}(E)+ \frac{\partial f_{R}(E)}{\partial\mu}\Delta\mu+\frac{\partial f_{R}(E)}{\partial T}\Delta T,
\end{equation}
in Eq. (\ref{current}), the response of the system is given by
\begin{eqnarray}\label{mut}
 \left( \begin{array}{c}
  J_{e} \\
  J_{q}
\end{array}
\right) &=&  \left( \begin{array}{cc}
              L_{11} &  L_{12} \\
              L_{21} & L_{22}
            \end{array} \right) \left( \begin{array}{cc}
              \Delta \mu/ eT \\
              \Delta T/T^2
            \end{array} \right),
\end{eqnarray}
where
%$ \Delta V=  \Delta \mu/e $ is the voltage drop across the molecular junction and
the Onsager coefficients $L_{11},L_{12},L_{21}$, and $L_{22}$ are given by
\begin{eqnarray}\label{onsager}
   L_{11} &=& \frac{T^2}{h}\int dE \,\tau(E)
\left[-\frac{\partial f(E)}{\partial E}\right], \nonumber\\
  L_{12} &=& \frac{Te}{h}\int dE \,\tau(E)
\left[-\frac{\partial f(E)}{\partial E}\right](E-\mu), \nonumber\\
  L_{22} &=& \frac{T}{h}\int dE \,\tau(E)
\left[-\frac{\partial f(E)}{\partial E}\right](E-\mu)^{2}, \nonumber\\
  L_{21} &=& L_{12}.
\label{eq:Lij}
\end{eqnarray}

The (isothermal) conductance $G$, defined as the electric current
under the application of the voltage $ \Delta \mu/e$ with no temperature gradient, is
\begin{equation}\label{econ}
    G = \frac{e J_{e}}{\Delta \mu} \bigg | _{\Delta T=0}=\frac{L_{11}}{T}.
\end{equation}
The thermal conductance $\Xi$, the heat current per unit temperature gradient for zero electric current, is
\begin{equation}\label{hcon}
    \Xi = \frac{J_{q}}{\Delta T} \bigg | _{J_{e}=0}=\frac{L_{11}L_{22}-L_{12}^{2}}{L_{11}T^2},
\end{equation}
and the Lorenz number $\mathfrak{L}$ is
\begin{equation}\label{lorentz}
    \mathfrak{L}=\frac{\Xi}{G T}.
\end{equation}
For a smooth function $\tau(E)$, the Sommerfeld expansion~\cite{asc} of the integrals
in (\ref{eq:Lij}) to lowest order in $k_B T/ E_F$, with
$E_F$ the Fermi energy, leads to the WF law:
\begin{equation}\label{lor}
\mathfrak{L}=\mathfrak{L}_0=\frac{\pi^2}{3}\left(\frac{k_B}{e}\right)^2.
\end{equation}
Note that to derive Eq. (\ref{lor}) from Eqs. (\ref{econ}) and (\ref{hcon}), the $L_{12}^{2}$ term has to be neglected, i.e., one needs $L_{11}L_{22} \gg L_{12}^{2}$ \cite{Beenakker}.

In the following sections we investigate in detail the dependence of the Lorenz number $\mathfrak{L}$ on the temperature $T$ and coupling strength $\gamma$ for a clean and disordered molecular wire.

\section{Clean Wire}
\label{sec:Clean Wire}

\begin{figure}[h!]
  \begin{center}
  \epsfig{file=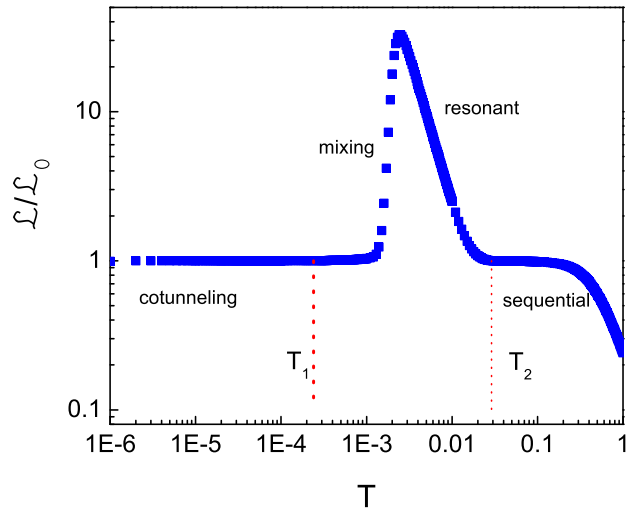,width=10cm}
  \end{center}
  \caption{Dependence of Lorenz number on temperature $T$ for a wire of length $N=110$ coupled to electrodes with strength $\gamma=10^{-4}$. Here $\mathfrak{L_{0}}=\pi^{2}/3$. Note that the WF law is satisfied in two regimes at low temperatures. All the parameters in this figure and through out the paper are scaled in units of hopping constant $t$. Also, we set $\hbar=e=k_B=1$, $t=1$ and $\mu=0$. For a typical wire of phenyl dithiol connected to Au electrodes, $T=1$ in our units corresponds to temperature of the order of $10^{4}K$. (See, e.g., Ref. \cite{mwpara}). } \label{te1}
\end{figure}

The transport mechanisms in the molecular system can be understood from the transmission function $\tau(E)$ of  the molecular wire. For a wire of $N$ sites, there are $N$ quantum states with discrete energies. The density of states and $\tau(E)$ in the
limit $\gamma\to 0$ consist of series of delta functions corresponding to these energies. While coupling to electrodes, electrons can enter or leave the wire and hence these delta peaks are broadened due to the finite life time of the electrons. If the coupling is very weak, then the densities of states remain as delta peaks broadened by a factor proportional to the coupling strength $\gamma$, whereas for a strong coupling all the peaks merge. Depending upon its energy relative to the energy spacing between the different levels in the wire ($\Delta E$), an electron can tunnel the junction mainly in three ways.

1) \emph{Sequential tunneling}: At temperature $T \gg \Delta E$,  the energy of the electron is very high and hence can tunnel across the junction in a sequential manner by spending a finite life time within the wire. Hence, the current in this regime is proportional to the coupling $\gamma$. 

2) \emph{Cotunneling}: This is a second order tunneling process occurring at temperature $T \ll \Delta E $.
In this regime, the current through the wire varies quadratically with the coupling strength $\gamma$. 

3) \emph{Resonant tunneling}: This occurs when the energy of an electron matches exactly one of the discrete energy levels in the wire.  Under these conditions, the electron is transmitted with unit probability and the current through the system increases sharply. Furthermore, at temperature $T \lesssim \Delta E$, a \textit{mixing} of the first-order and the second-order tunneling processes occurs.

To investigate the validity of the WF law, we study the variation of Lorenz number with temperature $T$. Figure \ref{te1} shows one example of our findings with a wire of $N=110$ sites. We found two plateaus where the law is satisfied exactly, i.e., when the ratio $\mathfrak{L/L_{0}}$ is $1$. Our numerical analysis shows that these two valid regimes correspond to sequential and cotunneling processes. Between these two plateaus, there is a region bounded by the temperatures $T_{1}$ and $T_{2}$ where,
upon lowering $T$, the Lorenz number increases initially due to resonance tunneling and thereafter decreases when  mixing of higher order tunneling occurs. Furthermore, at high temperatures the Lorenz number decreases quadratically.

\begin{figure}[h!]
  \begin{center}
  \epsfig{file=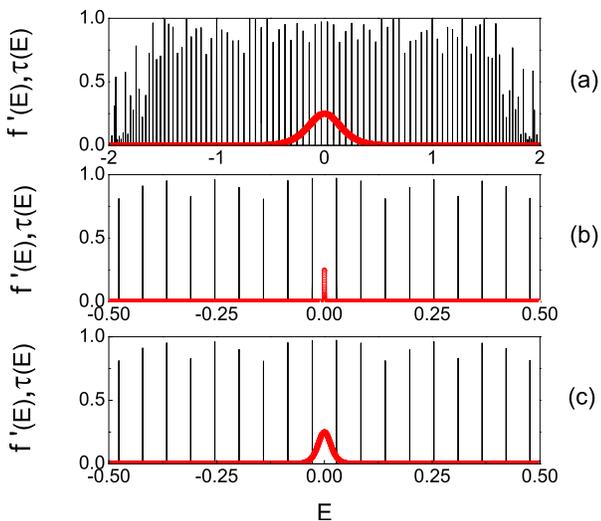,width=9cm}
  \end{center}
  \caption{Derivative of the Fermi distribution function $f'(E)=\frac{\partial f(E)}{\partial E}$ (thick curve) at temperatures (a) $T=0.1$, (b) $T=0.00001$, and (c) $T=0.01$. In the background, the transmission function $\tau(E)$ of a wire of $N=110$ sites is plotted as thin vertical lines. Note that $\tau(E)$ is a smooth function only in the sequential tunneling (a) and cotunneling (b) regimes.  } \label{te0}
\end{figure}

Validity regimes of the WF law can be understood~\cite{viol} from the energy integrals of the Onsager coefficients in Eq. (\ref{onsager}). In particular, when the derivative of the Fermi distribution function is sharp ($T\ll \Delta E$) or broad
($T\gg \Delta E$), the transmission function can
be considered a smooth function or its energy dependence can be averaged out,
respectively. Hence, the Sommerfeld expansion of the integrals leads to the WF law provided that $L_{11}L_{22} \gg L_{12}^{2}$ and $k_BT/E_F\ll 1$. At temperatures where the 
derivative of the Fermi distribution function is neither too sharp nor too broad,
the transmission peaks are not averaged out and violation of the WF law is expected. To illustrate this, the derivative of the Fermi distribution function (thick lines) in different tunneling regimes and the transmission function $\tau(E)$ (thin lines) of the investigated wire are plotted in Fig. \ref{te0}. Note that the conduction band width of our model is $[-2,2]$ and the mean level spacing $\Delta E\sim.03$.  As the electrochemical potential $\mu=0$, the Fermi distribution function is peaked around the energy $E=0$.  1)
In the \textit{sequential tunneling} regime (at temperature $T \gg \Delta E$) the derivative of the Fermi distribution function is broad [see Fig. 2(a)] so that peaks in the transmission function $\tau(E)$ are averaged out. 2) At temperatures in the \textit{cotunneling} regime ($T \ll \Delta E$), the derivative of the Fermi distribution function is sharp and hence the transmission function $\tau(E)$ can be considered as a
smooth, constant function on the scale $k_BT$ where the derivative of the
Fermi distribution function is significantly different from zero. This is clear form panel (b). 3) In the intermediate temperature regimes ($T \sim \Delta E$) where \textit{resonant tunneling} or \textit{mixing} occurs, the derivative is neither too sharp nor too broad and the different transmission peaks are not smoothened as shown in bottom panel of Fig. \ref{te0}.
Thus, the Sommerfeld expansion of the energy integrals in Eq. (\ref{onsager})
can be applied only in the sequential and cotunneling regimes, so that WF
law is valid only in these two regimes.
Note that for $\mu=0$, $L_{12}=0$ and hence we satisfy the relation $L_{11}L_{22} \gg L_{12}^{2}$. Deviation of the Lorenz number $\mathfrak{L}$ from the constant value $\mathfrak{L_{0}}$  at large temperatures in the sequential tunneling regime is justified as it follows from the analytical derivation of WF law that the Lorenz number is obtained only at low temperature $T \ll E _{F}$ where Sommerfeld expansion holds. The conduction band width for our model is $[-2,2]$ and hence the Fermi energy, $E_{F}$ is of the order of $2$ (see for instance ref.~\cite{asc}). Also, at high temperatures the conduction band width is exceeded, resulting in the quadratic decrease in Lorenz number $\mathfrak{L}$.

\begin{figure}[h!]
  \begin{center}
  \epsfig{file=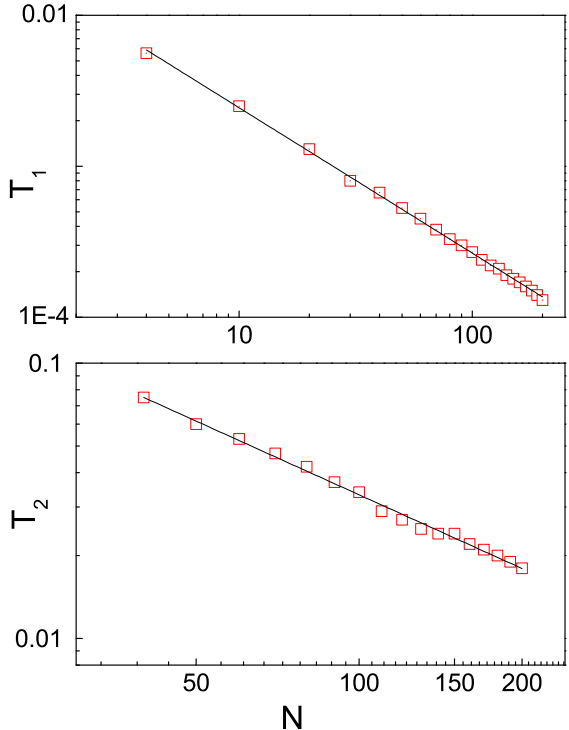,width=13cm}
  \end{center}
  \caption{Variation of characteristics temperatures $T_{1}$ and $T_{2}$ with the length $N$ of a molecular wire attached to electrodes with coupling strength $\gamma=10^{-4}$. Here $T_{1}$ is the highest temperature in the  cotunneling regime and $T_{2}$ is the lowest temperature in the sequential tunneling regime where the Lorenz number $\mathfrak{L}$=$\mathfrak{L_{0}}$.  Between $T_{1}$ and $T_{2}$, the WF law is not valid.  Note that a variation of the temperatures $\propto 1/N$ is obtained.} \label{te2}
\end{figure}

In order to clarify our statement that the WF law is satisfied only in the sequential and cotunneling regimes, we study the variation of $T_{1}$ and $T_{2}$ with the number of sites $N$. Note that these regimes correspond to cases where $T \gg \Delta E $ and $T \ll \Delta E $, respectively. The mean level spacing $\Delta E$ in the molecular wire decreases with the number of sites $N$ as $1/N$. Thus
dependences of temperatures $T_{1},T_{2}\propto 1/N$ are expected. This is indeed what we obtain in Fig. \ref{te2}. There, the top panel represents the highest temperature $T_{1}$ in the cotunneling regime where the WF law is satisfied. The bottom panel represents the lowest temperature $T_{2}$ in the sequential tunneling regime. The temperatures $T_{1}$ and $T_{2}$ are calculated such that the ratio $\mathfrak{L/L_{0}}$ is $1$ up to the third decimal point.

\begin{figure}[h!]
  \begin{center}
  \epsfig{file=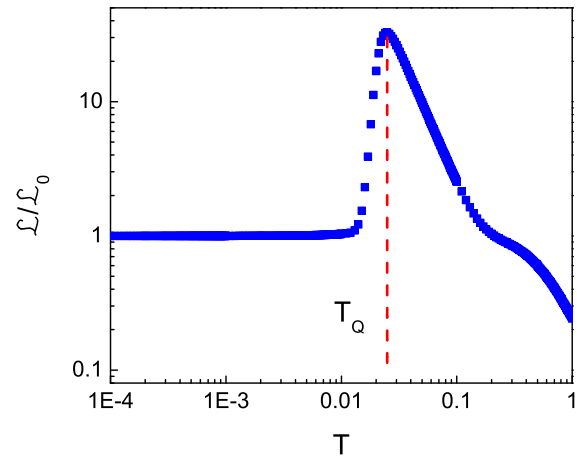,width=9cm}
  \end{center}
  \caption{Same as in Fig. \ref{te1} but for a wire of length $N=10$. Note that here the WF law is satisfied only in the cotunneling regime.} \label{te3}
\end{figure}

From Fig. \ref{te2}, it is clear that the temperature $T_{2}$ is $0.075$ for a chain of $40$ sites. The decrease $T_2\propto 1/N$ suggests that $T_{2}$ for $N=10$ should be $\simeq0.18$. In Fig. \ref{te3}, we have plotted the Lorenz number ratio as a function of temperature for $N=10$ sites. From the figure, it is clear that at $T=0.18$ the ratio $\mathfrak{L/L_{0}}$ is $0.953$. Also, the Lorenz number increases with
decreasing temperature $T$  without any saturation in the sequential tunneling regime. Indeed, numerical results showed that the plateau of constant Lorenz number seen in the sequential tunneling regime decreases with decreasing $N$ and is almost absent below $N=40$. This follows from the fact that for wires with $N \lesssim 40$, the mean level spacing $\Delta E$ is so large that the resonant tunneling occurs at temperatures higher than Fermi energy $E_{F}$.

\begin{figure}[h!]
  \begin{center}
  \epsfig{file=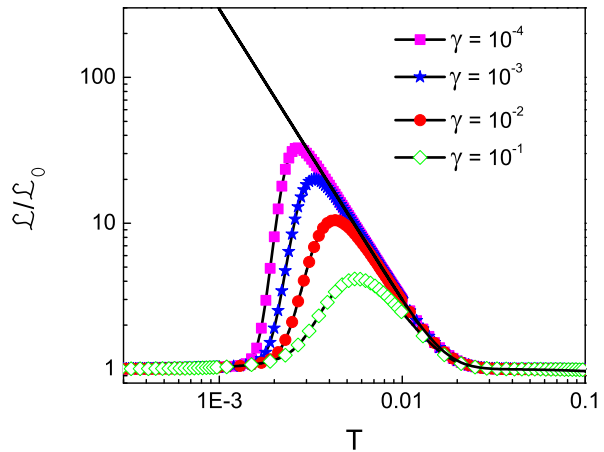,width=9cm}
  \end{center}
  \caption{Dependence of the Lorenz number on the electrode temperature $T$ for a wire of length $N=100$ with different coupling strengths $\gamma$. The straight line corresponds to the results obtained analytically for $\gamma\rightarrow 0$.} \label{te4}
\end{figure}

 So far we have discussed only the length dependence of the Lorenz number ratio for weak couplings to the electrodes. However, the strength of the coupling also plays an important role in modifying the transmission spectrum. Hence we have investigated the Lorenz number as a function of the coupling strength $\gamma$ in a wire of length $N=100$ in Fig. \ref{te4}. As shown in the figure, the large variation in Lorenz number is smoothed for stronger coupling. This is due to the fact that a strong coupling broadens the transmission function such that different resonance peaks overlap. Hence, it is difficult to observe the sharp increase in current due to the delta peaked transmission function.
\begin{figure}[h!]
  \begin{center}
  \epsfig{file=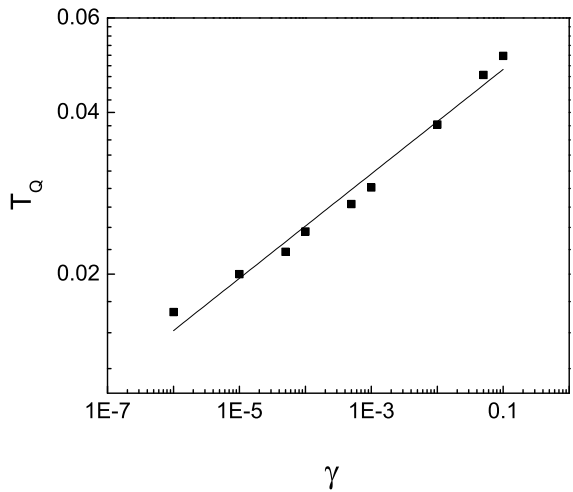,width=9cm}
  \end{center}
  \caption{Temperature $T_{Q}$ as a function of the coupling strength $\gamma$ for a wire of length $N=10$. Here $T_{Q}$ is the temperature at which the Lorenz number takes its maximum value.}
\label{te5}
\end{figure}

\begin{figure}[h!]
  \begin{center}
  \epsfig{file=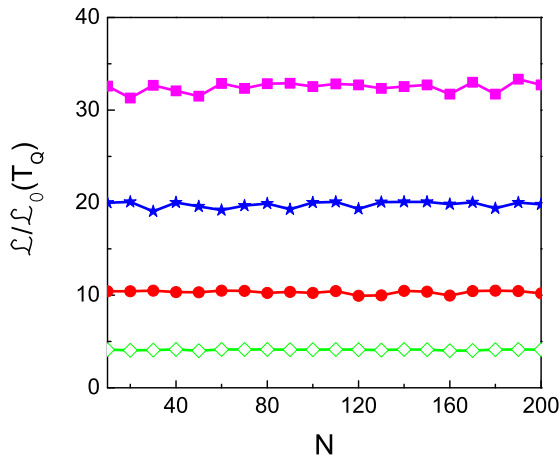,width=9cm}
  \end{center}
  \caption{Length dependence of the peak value of the Lorenz number with different coupling strengths $\gamma$. The lines from top to bottom correspond to $\gamma=10^{-4},10^{-3},10^{-2},$ and $10^{-1}$, respectively. For constant $\gamma$, the maximum value of the Lorenz number is independent of the length of the wire.} \label{te12}
\end{figure}

Our numerical analysis shows that the temperature $T_{Q}$ at which $\mathfrak{L}$ is maximum and below which the mixing regime (mixing of first-order and second-order tunneling process) occurs linearly increases with $\gamma$. This is illustrated in Fig. \ref{te5}, where $T_{Q}$ is plotted versus $\gamma$ for a wire of $10$ sites. The temperature $T_{Q}$ decreases from $0.051$ to $0.017$ as the coupling strength $\gamma$ is varied from $0.1$ to $10^{-6}$.
Also, $T_{Q}$ decreases $\propto 1/N$ with increasing length of the wire.

Another important observation is that for fixed coupling strengths $\gamma$ the magnitude of violation of the WF law is almost independent of the length of the wire. This is illustrated in Fig. \ref{te12}, where the value of $\mathfrak{L/L_{0}}$ at temperature $T_{Q}$ is plotted for different lengths of the wire. For a wire coupled to the electrode with $\gamma=0.1$, the maximum value obtained for the ratio of Lorenz number is around $4$. As the coupling strength is reduced to $10^{-4}$, the maximum value obtained increases to $32$. Indeed, our analytical calculations show that the Lorenz number diverges as $1/T^{2}$ for $\gamma\rightarrow 0$  (Details of the calculations are given in the Appendix.)  This perturbative (in $\gamma$) result is plotted as a straight line in Fig. \ref{te4}.

\begin{figure}[h!]
  \begin{center}
  \epsfig{file=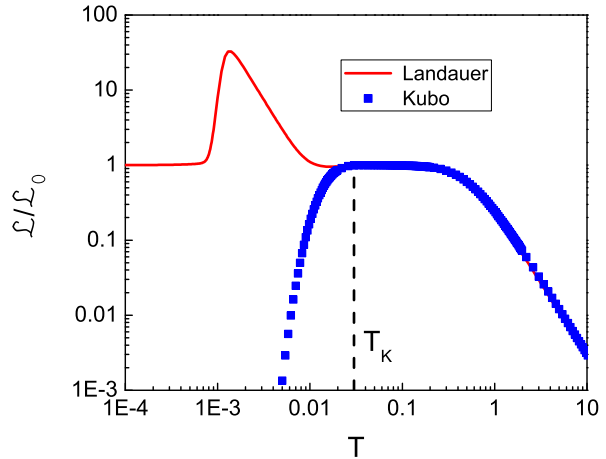,width=9cm}
  \end{center}
  \caption{ Comparison of the Landauer and Kubo formula for the Lorenz number of a wire of length $N=200$. The wire is attached to electrodes with a strength $\gamma=10^{-4}$. Note that the results of the Kubo formula coincide with those of the Landauer formula only above a temperature $T_{K}$.}\label{te6}
\end{figure}

We have also examined the narrow band limit of the electrodes with its density of states modeled as Lorentzian centered at zero energy with a width $\nu_{D}$. In the sequential tunneling regime, there is smoothening of the transmission peaks only with larger widths $\nu_{D}$ or longer wires. Hence, in general the WF law is not recovered in the sequential tunneling regime. However, for all values of $\nu_D$ the WF law
is found to be valid at low temperatures.

Another approach commonly used in investigating the transport properties of finite systems is the Kubo formalism. Here, details regarding the baths (in our case, electrodes) and coupling to the system (wire) are neglected and only the steady-state distribution of the system is used.  These assumptions are justified for investigating the linear response of an infinite system. However, formulas for finite systems are derived by extrapolating results of $N\rightarrow\infty$. The Lorenz number is computed in
terms of electrical and thermal conductivities, $\sigma$ and $\kappa$,
instead of the conductances $G$ and $\Xi$, i.e.
$\mathfrak{L}=\kappa/\sigma T$.
The Onsager coefficients are given by
 \begin{eqnarray} \label{kubo}
     L_{11}=e^2 T [D_{11}\delta(\omega) + \sigma_{11}(\omega)], \nonumber \\
   L_{12}=e T[D_{12}\delta(\omega) + \sigma_{12}(\omega)],  \nonumber \\
   L_{22}=T[D_{22}\delta(\omega) + \sigma_{22}(\omega)].
   \end{eqnarray}
 Here,
   \begin{eqnarray}
     D_{lm}=\frac{\pi \beta^{m}}{ZN}\sum_{\substack{i,k\\E_{i}=E_{k}}}e^{-\beta E_{i}}\langle i |j_{l}|k\rangle\langle k|j_{m}|i\rangle, \nonumber \\
      \sigma_{lm}(\omega)=\frac{\pi \beta^{m-1}}{ZN}\frac{1-e^{-\beta\omega}}{\omega}\sum_{\substack{i,k\\E_{i}\neq E_{k}}}e^{-\beta E_{i}} \nonumber \\ \times \langle i |j_{l}|k\rangle\langle k|j_{m}|i\rangle \delta(\omega-\Delta E).
   \end{eqnarray}
$E_{i}$ and $|i\rangle$ are the $i$th eigenenergy and eigenstate of the system, $\Delta E=E_{i}-E_{k}$; $\omega$ is the frequency, $\beta=1/k_BT$; $Z$ is the partition function; and $j_{1}$ and $j_{2}$ are the charge and heat currents. Currents $j_{1}$ and $j_{2}$ are calculated as $j_{1}=J_{e}$ and $j_{2}=J_{q}-\mu J_{e}$  where $J_{e[q]}=i\sum_{l=1}^{N-1}[h_{l-1},d_{e[q]}]$. $d_{e}$ is the number of electrons in the wire and  $d_{q}=h_{l-1}$ is the local system Hamiltonian. Note that $H_{W}=\sum_{l=1}^{N-1}h_{l}$, where $h_{l}=-t(c_{l}^{\dag}c_{l+1}+{\rm h.c.})$.

\begin{figure}[h!]
  \begin{center}
  \epsfig{file=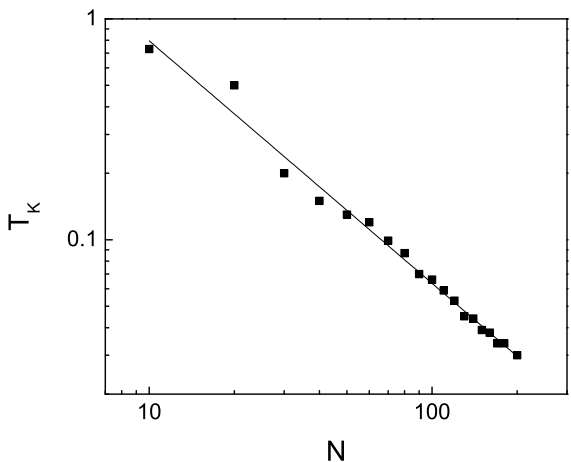,width=9cm}
  \end{center}
  \caption{Variation of the temperature $T_{K}$ with the length of the wire $N$. Here $T_{K}$ is the temperature at which the Kubo formula deviates from that of Landauer's. Note that $T_{K}$ varies $\propto1/N$ with the length of the wire $N$.}\label{te8}
\end{figure}

Figure \ref{te6} shows the ratio of the Lorenz number calculated using the above formula for our model with $N=200$ sites. For comparison, results obtained using the
Landauer-B\"uttiker formalism are also plotted. The results are the same until temperature $T_K=0.03$, below which the ratio computed using the Kubo formula deviates from that using the Landauer formula. We found that the temperature $T_{K}$ decreases  $\propto 1/N$ with an increase in the length $N$ of the wire. This can be understood as follows.  Eqs. (\ref{kubo}) are exact only for infinite systems for which the partition function $Z=\sum_{i}e^{-\beta E _{i}} \gg 1$. For small molecular wires, $Z$ can be large only for high temperature $T$. Figure \ref{te8} shows a plot of the temperature $T_{K}$ with the variation in length of the wire $N$. It is clear from Figs. \ref{te2} and \ref{te8} that the temperature $T_{2} <T_{K}$. This implies that the Kubo formula gives the results only for the sequential tunneling regime and hence can reproduce only one regime of the WF law even in arbitrarily long wires.

Finally, we have compared the results obtained by means of the Landauer-B\"uttiker
approach vs the Redfield QME.~\cite{saito,breuer,divkubo}
By construction, the QME is first-order perturbative in the coupling $\gamma$
and hence reproduces the results in the perturbative regime of the Landauer formula. This regime is bounded by the temperature $T_{Q}$ from below.  Since $T_1<T_Q$, the QME cannot reproduce the WF law
in the cotunneling regime.

\section{Disordered Wire}
\label{sec:Disordered Wire}
In this section, we discuss the validity of the WF law in a disordered wire. We model the disorder by introducing on-site energies $ \epsilon_{i}$ with randomness. The Hamiltonian of such a wire is

\begin{equation}\label{dw}
    {H}_{dW}=-t\sum_{i=1}^{N-1}(c_{i}^{\dag}c_{i+1}+{\rm h.c.})+\sum_{i=1}^{N} \epsilon_{i}c_{i}^{\dag}c_{i},
\end{equation}
where $ \epsilon_{i}$ are random numbers uniformly distributed in the interval $[-W,W]$. In one dimension, even for an arbitrary low disorder strength, the system becomes exponentially localized and exhibits insulating behavior.  \cite{loc1} Conductances of disordered wires decrease exponentially with the length of the wire as $G=G_{0}e^{-N/\xi}$, $\Xi=\Xi_{0}e^{-N/\xi}$, where $\xi$ is the localization length. Also, distributions of the conductances are log-normal parameterized solely by its mean value. However, these conclusions are true only at $T=0$ K. At any non-zero temperature, the exponential decrease is not apparent, as the electrons can hop from one localized state to another. \cite{tho}

\begin{figure}[h!]
  \begin{center}
  \epsfig{file=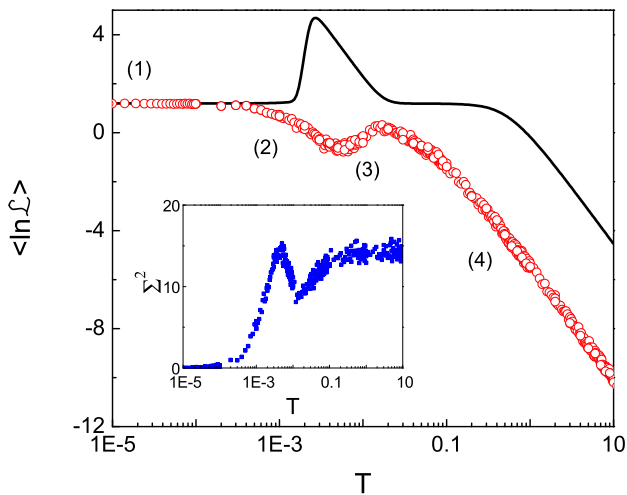,width=9cm}
  \end{center}
  \caption{Logarithmic average of the Lorenz number $\mathfrak{L}$ as a function of the electrode temperature $T$ for a wire of length $N=100$ with disorder strength $W=5$ and coupling strength $\gamma=10^{-4}$. The straight line corresponds to the case of a clean wire. Here regimes (1), (2), (3) and (4) corresponds to cotunneling, mixing, resonant tunneling, and sequential tunneling, respectively.  Note that only in the cotunneling regime are the results the same for both wires. Inset: Temperature dependence of the variance $\sum^{2}$ of $\mathfrak{L}$.  }\label{te9}
\end{figure}

The localization length $\xi$ is maximum at the band center and decreases towards the band edge. \cite{locd} At each energy $E$,  the localization length is related to the transmission function as $\xi(E)^{-1}= -\ln \tau(E)/2 N$ for a wire of $N$ sites. \cite{locdt}  We restrict our analysis to wires with $\xi \ll N$ so that the system is insulating. For this we consider a wire of length $N=100$ with disorder strength $W=5$. Our calculations show that the localization length of the wire is $\xi < 1$ at all energies $E$.
As the conductivities exhibit giant fluctuations for different samples, we take logarithmic averages. The dependence of the logarithm of the Lorenz number  $\langle \mathrm{ln}\mathfrak{L}\rangle$ on the temperature $T$ for this model is depicted in Fig. \ref{te9}. The values are obtained by taking  the average over $1500$ disorder realizations. The logarithm of the Lorenz number $\mathfrak{L_{0}}=\pi^2/3$ is equal to $1.1908$ (in units where $e=k_B=1$). From the figure, it is clear that the plateau of constant Lorenz number $\mathfrak{L_{0}}$ is recovered at low temperatures. This corresponds to the cotunneling regime  indicated by (1) in the figure. As the temperature is increased, the mixing regime [region (2) in Fig. \ref{te9}] is reached, where the Lorenz number decreases with increases in temperature. The decrease is apparent till the temperature where the resonant tunneling occurs [regime (3)]. Further increases in temperature increases the Lorenz number until the sequential tunneling regime. Finally, in the sequential tunneling regime as shown by (4) in the figure, the Lorenz number decreases quadratically with temperature.

To better understand the differences in the variation of Lorenz number with temperature for disordered and nondisordered wires, we have plotted the variation of the logarithm of the Lorenz number $\mathfrak{L}$ for a clean wire as a straight line in Fig. \ref{te9}. Our findings in this regard are summarized as follows. (a) The Lorenz number for the disordered wire is always equal to or less than that of the clean wire at all temperatures. (b) There is no saturation plateau of constant Lorenz number in the sequential tunneling regime. This is in contrast to the results in Ref. \cite{viol}, where it was pointed out that the WF law is violated in the resonant tunneling regime and  is valid in the sequential tunneling regime.  We note that due to the finite size of our wire, there is no self-averaging of the transmission peaks over the window $k_{B}T$.  (c) The temperature $T_{1}$ at which the Lorenz number $\mathfrak{L_{0}}$ is recovered is shifted to lower temperature. This follow from disorder-induced energy fluctuations, so that there
exist samples for which the spacing between the Fermi energy and the nearest
peak of $\tau(E)$ is much smaller than the mean level spacing $\Delta E$.
In such instances, the Sommerfeld expansion substantiating WF law is valid only
at lower temperatures than in the clean case.

So far we have focused only on the mean value of the Lorenz number. However, in a disordered system  with a localization length much smaller than the system size, i.e., $\xi \ll N$, fluctuations can be as large as the average value. Under these conditions, only a statistical distribution provides meaningful information about the system properties.
It is well known that in these highly localized systems, the conductance distribution is log-normal i.e., ${\ln}G$ (or ${\ln}\Xi$) follows a normal distribution at $T=0$ K. The finite-temperature transport properties of the disordered wire can be understood in terms of the distribution statistics of the transmission function $\tau(E)$. This is clear from  Eq. (\ref{onsager}) where the Onsager coefficients are expressed in terms of the transmission function $\tau(E)$. However, to derive the distribution of the Lorenz number using Eqs. (\ref{onsager})-(\ref{lorentz}) is beyond the scope of the current study. Hence, in the latter part of this section, we analyze numerically, in detail, the distribution followed by the Lorenz number in the different tunneling regimes.

\begin{figure}
  \begin{center}
  \epsfig{file=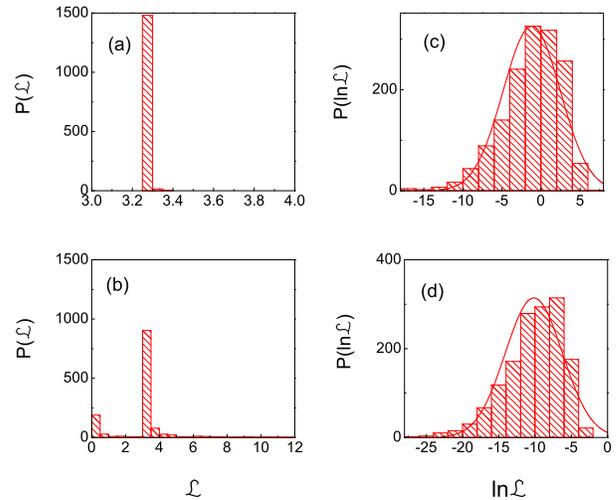,width=9cm}
  \end{center}
  \caption{Distribution of Lorenz number $\mathfrak{L}$ at different temperature regimes for a wire of length $N=100$ with disorder strength $W=5$. Different panels correspond to temperatures (a) $T=0.00001$ in the cotunneling regime; (b) $T=0.001$ in the mixing regime; (c) $T=0.01$ in the resonant tunneling regime; (d) $T=10$ in the sequential tunneling regime. Note that in (c) and (d), the logarithm of the Lorenz number is plotted. }\label{te10}
\end{figure}

\subsection{Cotunneling regime}
It is clear from Fig. \ref{te9} that in this regime we reproduce the WF law. From our numerical analysis we found that fluctuations of the conductances are almost twice as that of the average value. This has been advocated as the evidence for insulating behavior in one dimensional system \cite{var2mean} and thus we ensure that disordered wire is indeed localized. We find that the fluctuations of both conductances are almost perfectly correlated and hence the Lorenz number $\mathfrak{L_{0}}$ with almost zero variance is obtained. This is clear from Fig. \ref{te10} (a), where the distribution at temperature $T=0.00001$ is shown. A delta peak around the value $\pi^{2}/3=3.2898$ is obtained at all temperatures in this regime.

\subsection{ Mixing regime}
 In this regime we found that the logarithm of conductances still follows a normal-like distribution, with variance almost twice that of the average. However, the skewness is non-zero. Moreover, the fluctuations of both conductances are not perfectly correlated and therefore the constant value of the Lorenz number $\mathfrak{L_{0}}$ is not recovered. Interestingly, we obtain a bimodal distribution for the Lorenz number. This is shown in  Fig. \ref{te10} (b) for temperature $T=0.001$. One of the two peaks corresponds to the value $\pi^{2}/3$, while the other corresponds to a value near to $0$. The peak around $\pi^{2}/3$ is largely populated at low temperatures. Upon increasing the temperature, this peak is reduced, while the peak with a small Lorenz number is populated. This continues till the resonant tunneling regime is reached where there is only one peak. Indeed, our results suggest that the two peaks correspond to first-and second-order tunneling processes. For disordered wires, second-order tunneling favors the WF law with Lorenz number $\mathfrak{L_{0}}$, whereas first-order tunneling favors a small value of the Lorenz number. Also, the variance of the Lorenz number in this regime increases with increases in temperature until the two peaks are equally populated and decreases thereafter.

\subsection{Resonant tunneling regime}
 Here, similarly to the mixing regime case, the logarithm of conductances has a skew normal distribution (a normal distribution but with non-zero skewness).  In contrast to the mixing regime, the variances of the conductances decrease more rapidly with temperature compared to their mean values and are almost of the order of the mean values. The distribution of the Lorenz number in this regime is always peaked near zero value with long tails toward large values and can be approximated to a log-normal distribution. Hence, we plotted the distribution of the logarithm of $\mathfrak{L}$. A typical distribution in this regime at temperature $T=0.01$ is shown in Fig. \ref{te10} (c) and is in accordance with our expectation. Furthermore, it is shown that the variance in this regime increasing with increase in temperature.
\begin{figure}
  \begin{center}
  \epsfig{file=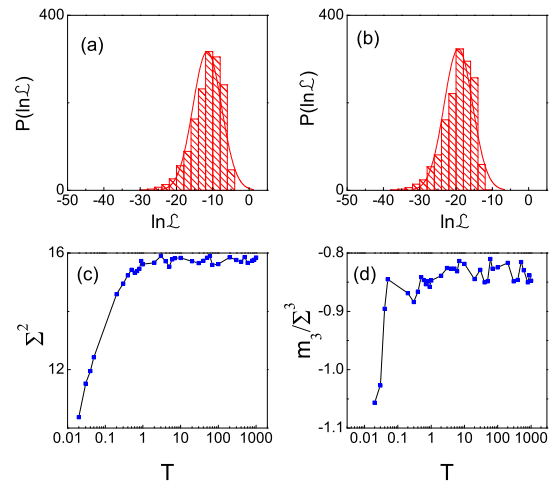,width=9cm}
  \end{center}
  \caption{Top: Distribution of the Lorenz number for a wire of length $N=100$ with disorder strengths $W=5$ in the sequential tunneling regime at temperatures (a) $T=20$ and (b) $T=1000$.  (c) Variance ($\Sigma^{2}$) and (d)  skewness ($m_{3}/\Sigma^{3}$) of the Lorenz number distribution at different temperatures in the sequential tunneling regime.}\label{te11}
\end{figure}

\subsection{Sequential tunneling regime}
Our results in this regime indicate that the logarithm of the conductances has a distribution similar to that of the resonant tunneling regime. However, the variance initially decreases with increasing temperature and thereafter saturates.
The distribution of the Lorenz number is still peaked around a very small value. Hence, similarly to the case of the  resonant tunneling regime, we have plotted the logarithmic distribution in Fig. \ref{te10}(d). Here, a skew normal distribution of increasing variance and decreasing skewness
is obtained as the temperature is initially increased. However, after the initial change, both the variance and the skewness saturate to a constant value as the fluctuations of the conductances also saturate. Thus, a distribution invariant with temperature emerges and is demonstrated in  Figs. \ref{te11}(a) and (b), which correspond to temperatures $T=20$ and $1000$, respectively. In order to better clarify the emergence of the invariant distribution, we have plotted the variance ($\Sigma^{2}$) and the skewness ($m_{3}/\Sigma^{3}$, where $m_{3}$ is the third moment about the mean) of the distributions in Figs. \ref{te11}(c) and (d). It is clear from the figure that the variance saturates around the value $15$, whereas the skewness fluctuates around $-0.85$ at high temperatures, indicating a temperature-invariant
distribution.

\section{Conclusions}
\label{sec:Conclusions}
Using the Landauer-B\"uttiker formalism, we have investigated the validity of the WF law in finite-sized molecular wires with and without disorder. For a clean system, we found that the validity regimes of the WF law depend on how an electron tunnels across the wire. In particular, the Lorenz number $\mathfrak{L_{0}}=(\pi^{2}/3)(k_B/e)^2$ is obtained in the cotunneling and the sequential tunneling regime as long as the temperature in these regimes is much lower than the Fermi energy $E_{F}$. For wires of length $N \lesssim 40$, resonant tunneling occurs for temperatures higher than the Fermi energy $E_{F}$ and hence the WF law is valid only in the cotunneling regime. Following the studies of Vavilov and Stone in Ref. \cite{viol}, the violation of the law in the different tunneling regimes is explained in terms of the energy integrals giving the electrical and thermal conductances. We have further compared our results with the standard Kubo formula and with the Redfield QME and found that the two approaches diverge from the Landauer formula at particular temperatures $T_{K}$ and $T_Q$, which decrease with an increase in the length of the wire $N$. The temperatures $T_{K}$ and $T_Q$ are always higher than the temperature at which cotunneling occurs and hence it follows from our results that the Kubo formula and QME will differ from the Landauer-B\"uttiker formalism even in the limit of infinite length of the wire.

Furthermore, we have explored an exponentially localized disordered wire  using the
Landauer-B\"uttiker formalism. Here, even for wires of length $N=100$, the WF law is valid only at very low temperatures corresponding to the cotunneling regime. Moreover, the Lorenz number shows typical distributions at different temperatures corresponding to different tunneling processes.  A delta distribution peaked around the value $\mathfrak{L_{0}}$ is obtained in the cotunneling regime, while a bimodal distribution is obtained in the regime where the mixing of first- and second-order tunneling processes occurs. The logarithm of the Lorenz number shows a skew normal distribution in the resonant and sequential tunneling regimes. In particular, we found that a distribution with constant variance and skewness emerges in the high-temperature regime. We infer from our results that first-order tunneling favors a small Lorenz number whereas second-order tunneling favors the universal value $\mathfrak{L_{0}}$ in a disordered wire.

Finally, we point out that we have not addressed the effects of interaction between the electrons in our model. Commonly used approaches to investigating the transport properties of interacting finite-sized systems are the standard Kubo formalism and QME. We have discussed numerically in detail the failure of the Kubo formalism to obtain the WF law in finite-sized systems. Furthermore, it follows from our numerical analysis that the standard QME approach also fails. The standard QME is derived by taking second-order perturbative expansion of tunneling amplitudes $t_{kj}$ and is only linear in coupling strength $\gamma$. Hence, the current calculated using this formalism is always of first order in $\gamma$. However, in the cotunneling regime the current varies quadratically with $\gamma$.  A fourth-order perturbative expansion of tunneling amplitudes in the QME indeed explains this regime. \cite{4order} Thus, we note that to investigate the Lorenz number in finite-sized strongly interacting systems, the standard QME has to be extended to include terms of $t^{4}_{kj}$ .

\section*{Acknowledgements}

We acknowledge support by the
MIUR-PRIN 2008 and by Regione Lombardia.

\appendix

\section{ANALYTICAL DERIVATION}

In this section, we analytically derive the electric ($J_{e}$) and thermal ($J_{q}$) currents for a wire weakly coupled to the electrodes ($\gamma\rightarrow 0$).

From the definition of the single-particle Green's function operator $G_{s}$, it follows that $G_{s}$ is the inverse of the matrix
\begin{eqnarray}\label{mut1}
    \left( \begin{array}{ccccc}
              E-i\frac{\gamma}{2} & -1 & . & . & . \\
              -1 & E & . & . & . \\
                0 & . & . & . & . \\
                . & . & . & E & -1 \\
                . & . & . & -1 & E+i\frac{\gamma}{2}
            \end{array} \right) .
\end{eqnarray}
Thus, $\langle 1|G_{s}|N\rangle$ is the $(1,N)$ element of the inverse of the above matrix.

Consider a wire with N=1 coupled to both the left and the right reservoirs.
Here, $\langle 1|G_{s}|N\rangle=1/(E-i{\gamma})$ and hence
%{\bf I think there is no - sign in the integral below, please check}
\begin{equation}\label{jn1}
     J_{e} = \frac{e \gamma^{2}}{2\pi\hbar}\int dE \frac{1}{E^{2}+\gamma^{2}}[f_{L}(E)-f_{R}(E)].
\end{equation}
Apart from the constant factor, the first integral is of the form
\begin{equation}\label{i1}
    I=\int dE \frac{1}{E^{2}+\gamma^{2}}f(E)\equiv \int dE g(E),
\end{equation}
which can be evaluated using the residue theorem. The first part of the integrand has two poles, namely, $E_{\pm}=\pm i \gamma$, whereas the second part has an infinite number of poles, namely, Mastubara frequencies at $E_{n}=\mu+\frac{i}{\beta}(2n+1)\pi$ with $n\in\mathbb{Z}$ and $\beta=1/T$ (we set $k_B=1$).

To evaluate the integral, we consider two distinct contours, one in the upper half-plane $(C_{1})$ and the other in the lower half-plane $(C_{2})$. Each contour runs from $-R$ to $R$ on the real axis and then comes back, following a semicircle of radius $R$. Thus, the integration domain is the sum of the following two parts:
\begin{eqnarray}\label{int1}
\int_{C_{1}}=\int_{-R}^{R} + \int_{C_{R}} \quad; \quad \int_{C_{2}}=\int_{-R}^{R} + \int_{C'_{R}}
\end{eqnarray}
where $C_{R}$ ($C'_{R}$) denotes the upper (lower) semicircle. Since the integrand function satisfies the "big circle lemma" (Jordan's lemma), the contribution of the integrals on the semicircles is $0$ when $R\rightarrow\infty$ and thus only the contribution from the real axis integration survives. Applying the residue theorem, we get
\begin{eqnarray}\label{int2}
  I&=&\frac{1}{2}\,2 \pi i [\mathrm{res} \, g(E_{+})-\mathrm{res}\, g(E_{-}) \nonumber \\ &+&\sum_{n=0}^{\infty}\mathrm{res}\,g(E_{n}) -\sum_{n=-1}^{-\infty}\mathrm{res}\,g(E_{n})].
\end{eqnarray}
The factor $1/2$ follows from the fact that integrating on two contours, real axis integration is encountered twice. Now the first two terms in Eq. (\ref{int2}) are
\begin{eqnarray}
  \mathrm{res} \,g(E_{+}) &=&  \lim_{E\rightarrow E_{+}}g(E)(E-E_{+}) \nonumber \\
&=& \lim_{E\rightarrow E_{+}}\frac{f(E)(E-E_{+})}{(E-E_{+})(E-E_{-})} \nonumber\\
&=&\frac{f(E_{+})}{(E_{+}-E_{-})},
   \end{eqnarray}
and
\begin{eqnarray}
  \mathrm{res} \,g(E_{-}) &=&  \lim_{E\rightarrow E_{-}}g(E)(E-E_{-}) \nonumber \\
 &=&-\frac{f(E_{-})}{(E_{+}-E_{-})}.
   \end{eqnarray}
Thus, their difference is
\begin{eqnarray}
 \mathrm{res} \,g(E_{+})-\mathrm{res} \,g(E_{-})&=&\frac{1}{(E_{+}-E_{-})}[f(E_{+})-f(E_{-})] \nonumber \\
&=&\frac{2\mathrm{Re}[f(E_{+})]}{(E_{+}-E_{-})} \nonumber \\
&=&\frac{1}{i\gamma}\mathrm{Re}([f(i\gamma)].
\end{eqnarray}
The residue corresponding to the $n$th Mastubara frequency $E_{n}$ is
\begin{eqnarray}
 \mathrm{res} \,g(E_{n})&=&\frac{1}{E^{2}+\gamma^{2}}\frac{1}{(\exp[\beta(E-\mu)]+1)'|_{E=E_{n}}} \nonumber \\
&=&\frac{1}{E^{2}+\gamma^{2}}\frac{1}{\beta\underbrace{(\exp[\beta(E-\mu)]+1)}_{=-1}}
\nonumber \\ &=&-\frac{1}{\beta}\frac{1}{E^{2}+\gamma^{2}}.
\end{eqnarray}

 Here we point out that
\begin{eqnarray}
 E_{-n-1}&=&\mu+\frac{1}{\beta}(-2(n+1)+1)i\pi \nonumber \\
&=&\mu+\frac{1}{\beta}(-2n-1)i\pi=E_{n}^{*}.
\end{eqnarray}
Also,
\begin{eqnarray}
 \sum_{n=-1}^{-\infty}\mathrm{res}\,g(E_{n})&=& \sum_{m=0}^{\infty}\mathrm{res}\,g(E_{-m-1}) \quad (n=-m-1) \nonumber\\
&=&\sum_{n=0}^{\infty}\mathrm{res}\,g(E_{n}^{*}).
\end{eqnarray}
Therefore, the difference between the residues is
\begin{eqnarray}
  \mathrm{res}\,g(E_{n})-\mathrm{res}\,g(E_{-n-1})&=&-\frac{1}{\beta}
\left[\frac{1}{E_{n}^{2}+\gamma^{2}}
-\frac{1}{E_{-n-1}^{2}+\gamma^{2}}\right] \nonumber\\
&=&-\frac{i}{\beta}\,2\,\mathrm{Im}\left[\frac{1}{E_{n}^{2}+\gamma^{2}}\right].
\end{eqnarray}
Substituting all these results in Eq. (\ref{jn1}) we get,
\begin{eqnarray}
  J_{e} &=& \frac{e\gamma^{2}}{2\pi\hbar}\left\{\frac{\pi}{\gamma}\mathrm{Re}[f_{L}(i\gamma)]-
\mathrm{Re}[f_{R}(i\gamma)]\right.
   \nonumber \\ && \left. +\frac{2\pi}{\beta_{L}}\sum_{n=0}^{\infty}\mathrm{Im}\left(\frac{1}{E_{nL}^{2}+\gamma^{2}}\right)\right. \nonumber \\ && \left.
 -\frac{2\pi}{\beta_{R}}\sum_{n=0}^{\infty}\mathrm{Im}\left(\frac{1}{E_{nR}^{2}+\gamma^{2}}\right)\right\},
\end{eqnarray}
where $E_{nL,R}=\mu_{L,R}+\beta_{L,R}^{-1}(2n+1)i \pi$ with $n\in\mathbb{Z}$.
In the limit of weak coupling, i.e., $\gamma \ll T$, the above equation reduces to
\begin{equation}\label{j1}
    J_{e} \simeq \frac{e\gamma}{2\hbar}[f_{L}(E=0)-f_{R}(E=0)].
\end{equation}
Note that in the last formula $E=0$ should be substituted by the dot's energy
$E=\epsilon_0$ in the case of the single-dot Hamiltonian
$H_W= \epsilon_0 c^\dagger c$.

For a wire with $N=2$,
\begin{equation}\label{g2}
    \langle 1| G_{s} |N\rangle=\frac{1}{(E-E_{1}+i\frac{\gamma}{2})(E-E_{2}+i\frac{\gamma}{2})},
\end{equation}
where $E_{j}=2\cos\{[\pi/(N+1)]j\}$ $(j=1,2)$. Thus, the current $J_{e}$ is given by
\begin{equation}\label{n1}
     J_{e} = \frac{e \gamma^{2}}{2\pi\hbar}\int dE \frac{f_{L}(E)-f_{R}(E)}{\left|(E-E_{1}+i\frac{\gamma}{2})(E-E_{2}+i\frac{\gamma}{2})\right|^2}.
\end{equation}
Following the same steps for $N=1$, we obtain
\begin{eqnarray}
  J_{e} &=& \frac{\gamma^{2}}{2\pi\hbar}\frac{1}{2} \,2 \pi i \left\{\frac{2\mathrm{Re}[f_{L}(E_{1}+i\frac{\gamma}{2})-f_{R}(E_{1}+i\frac{\gamma}{2})]}{i\gamma(E_{1}-E_{2}+i\gamma)(E_{1}-E_{2})}\right.
\nonumber \\ && \left. +\frac{2\mathrm{Re}[f_{L}(E_{2}+i\frac{\gamma}{2})-f_{R}(E_{2}+i\frac{\gamma}{2})]}{i\gamma(E_{2}-E_{1}+i\gamma)(E_{2}-E_{1})}\right\}.
\end{eqnarray}
In the limit of $\gamma\rightarrow 0$, the current $J_{e}$ is
\begin{eqnarray}
  J_{e} &\simeq & \frac{e \gamma}{\hbar} \left[\frac{f_{L}(E_{1})-f_{R}(E_{1})}{(E_{1}-E_{2})^{2}}\right. \nonumber \\
&&+ \left.\frac{f_{L}(E_{2})-f_{R}(E_{2})}{(E_{2}-E_{1})^{2}}\right].
\end{eqnarray}

Similarly, for a wire of length $N$ we get
\begin{equation}\label{eq1}
   J_{e}\simeq \frac{e \gamma}{\hbar} \sum_{k=1}^{N}\frac{f_{L}(E_{k})-f_{R}(E_{k})}{\prod_{j\neq k}(E_{k}-E_{j})^{2}},
\end{equation}
for the electric current and
\begin{equation}\label{eq2}
   J_{q}\simeq \frac{\gamma}{\hbar} \sum_{k=1}^{N}\frac{(E_{k}-\mu)[f_{L}(E_{k})-f_{R}(E_{k})]}{\prod_{j\neq k}(E_{k}-E_{j})^{2}},
\end{equation}
for the thermal current, with $E_{j}=2\cos\{[\pi/(N+1)]j\}$ $(j=1,...,N)$.
Using the Taylor expansion in Eq. (\ref{tf2}), the difference of the Fermi functions reads,
\begin{eqnarray}\label{f1}
  f_{L}(E_{k})-f_{R}(E_{k})
  &=& -\frac{e^{\beta (E_{k}-\mu)}}{(e^{\beta(E_{k}-\mu)}+1)^{2}}[(E_{k}-\mu)\frac{\Delta T}{T^{2}}+\frac{\Delta \mu}{T}] \nonumber \\
   &=&-\frac{1}{4\cosh^{2}[\frac{\beta(E_{k}-\mu)}{2}]}\nonumber \\
   &&\times[(E_{k}-\mu)\frac{\Delta T}{T^{2}}+\frac{\Delta\mu}{T}].
\end{eqnarray}
The above expressions are used in Eqs. (\ref{econ}), (\ref{hcon}), and (\ref{Lorenz}) to calculate the conductances $G$ and $\Xi$ and the Lorenz number $\mathfrak{L}$ in Fig. \ref{te4}. It is clear from the above analytical calculations that the conductances $G\propto 1/T$ and $\Xi \propto 1/T^{2}$ and thus the Lorenz number $\mathfrak{L} = \frac{\Xi}{G T} \propto 1/T^{2}$.


\begin{thebibliography}{100}

  \bibitem{te} G. Mahan, B. Sales, and J. Sharp, Phys. Today {\bf 50}, 42 (1997).

    \bibitem{Majumdar} A. Majumdar, Science {\bf 303}, 777 (2004).

\bibitem{dresselhaus}
M. S. Dresselhaus, G. Chen, M. Y. Tang, R. G. Yang, H. Lee,
D. Z. Wang, Z. F. Ren, J. -P. Fleurial, and P. Gogna,
Adv. Mater. \textbf{19}, 1043 (2007).

\bibitem{snyder}
G. J. Snyder and E. S. Toberer,
Nature Mater. \textbf{7}, 105 (2008).


\bibitem{shakuori}
A. Shakouri,
Annu. Rev. Mater. Res. {\bf 41}, 399 (2011).

\bibitem{BC11}
G. Benenti and G. Casati,
Phil. Trans. R. Soc. A {\bf 369}, 466 (2011).

    \bibitem{ztone} L. D. Hicks and M. S. Dresselhaus, Phys. Rev. B {\bf 47} (R), 16631 (1993).
\bibitem{heremans}
J. P. Heremans, V. Jovovic, E. S. Toberer, A. Saramat, K. Kurosaki, A. Charoenphakdee, S. Yamanaka, and G. J. Snyder, Science
{\bf 321}, 554 (2008)􏰒.
\bibitem{datta}
R. Kim, S. Datta, and M .S. Lundstrom, J. Appl. Phys. {\bf 105},
034506 (2009)􏰒.
\bibitem{linke}
N. Nakpathomkun, H. Q. Xu, and H. Linke,
Phys. Rev. B {\bf 82}, 235428 (2010).

       \bibitem{WF} G. Weidemann and R. Franz, Ann. Phys. {\bf 89}, 497 (1853).
        \bibitem{asc}  N. W. Ashcroft and N. D. Mermin, {\it Solid State Physics} (Saunders, Philadelphia, 1976).
        \bibitem{WFdisnon} G. V. Chester and A. Thellung, Proc. Phys. Soc. {\bf 77}, 1005 (1961).
       \bibitem{WFint1} C. L. Kane and M. P. A. Fisher,  Phys. Rev.
       Lett. {\bf 76}, 3192 (1996).
         \bibitem{WFint3}M. -R. Li and E. Orignac, Europhys. Lett. {\bf 60},
     432 (2002).
      \bibitem{WFint4}B. Dora, Phys. Rev. B {\bf 74}, 161101(R) (2006).
        \bibitem{WFint2}A. Garg, D. Rasch, E. Shimshoni, and A. Rosch, Phys. Rev.
       Lett. {\bf 103}, 096402 (2009).
      \bibitem{WFint5}N. Wakeham, A. F. Bangura, X. Xu, J. -F. Mercure, M. Greenblatt, and N. E. Hussey, Nature Commun. {\bf 2}, 396 (2011).
          \bibitem{viol} M. G. Vavilov and A. D. Stone, Phys. Rev. B {\bf 72}, 205107 (2005).
     \bibitem{Sfinite} M. M. Zemlji\v{c},  and P. Prelov\v{s}ek, Phys. Rev. B {\bf 71}, 085110 (2005).
    \bibitem{Lshastry} M. R. Peterson, S. Mukerjee, B. S. Shastry, and J. O. Haerter, Phys. Rev. B {\bf 76}, 125110 (2007).
    \bibitem{divkubo} J. Wu and M. Berciu, Europhys. Lett. {\bf 92},
     30003 (2010).
    \bibitem{Anderson} P. W. Anderson, Phys. Rev. {\bf 109}, 1492 (1958).
    \bibitem{conduc} C. W. J. Beenakker, Rev. Mod. Phys. {\bf 69}, 731 (1997).
    \bibitem{expdec} E. Abrahams, P. W. Anderson, D. C. Licciardello, and T. V. Ramakrishnan, Phys. Rev.
       Lett. {\bf 42}, 673 (1979).
    \bibitem{pichard} J. -L. Pichard, N. Zanon, Y. Imry,
and A. D. Stone, J. Phys. (Paris) {\bf 51}, 587 (1990).
    \bibitem{Landauer} S. Datta, {\it Electronic Transport in Mesoscopic Systems} (Cambridge University Press, Cambridge, 1995).
%    \bibitem{meta} M. E. Gerhenson, Yu. B. Khavin, A. G. Mikhalchuk, H. M. Bozier, and A. L. Bogdanov, \prl {\bf 79}, 725 (1997).
\bibitem{Beenakker} H. van Houten, L. W. Molenkamp, C. W. J. Beenakker, and C. T. Foxon, Semicond. Sci. Technol. {\bf 7}, B215 (1992).
    \bibitem{mwpara} J. W. Lawson and C. W. Bauschlicher, Phys. Rev. B. {\bf 74}, 125401 (2006).
\bibitem{saito}
K. Saito, S. Takesue, and S. Miyashita, Phys. Rev. E {\bf 61}, 2397 (2000).
\bibitem{breuer}
H. Wichterich, M. J. Henrich, H. -P. Breuer, J. Gemmer, and M. Michel,
Phys. Rev. E {\bf 76}, 031115 (2007).
    \bibitem{loc1} N. F. Mott and W. D. Twose, Adv. Phys. {\bf 10}, 107 (1961).
    \bibitem{tho} D. J. Thouless, Phys. Rev.
       Lett. {\bf 39}, 1167 (1977).
       \bibitem{locdt} P. W. Anderson, D. J. Thouless, E. Abrahams, and D. S. Fisher, Phys. Rev. B {\bf 22}, 3519 (1980).
       \bibitem{locd} B. Kramer and A. MacKinnon, Rep. Prog. Phys. {\bf 56}, 1469 (1993).
    \bibitem{var2mean} A. Abrikosov, Solid State Commun. {\bf 37} 997, (1980); N. Kumar, Phys. Rev. B {\bf 31}, 5513 (1985); P. Mello, J. Math. Phys. {\bf 27}, 2876 (1986).
        \bibitem{4order} F. Elste and C. Timm, Phys. Rev. B   {\bf 75}, 195341 (2007).
        %K. M. Slevin and J. B. Pendry, J. Phys. Condens. Matter {\bf 2} 2821 (1990).

        \end{thebibliography}
        \end{document}